\documentclass[9pt, journal]{IEEEtran}

\IEEEoverridecommandlockouts

\usepackage{graphicx,amsmath,amssymb,remark,color}
\newcommand{\eproof}{\hfill\rule{2mm}{2mm} \medskip}

\usepackage{cite,epsfig}
\newcommand{\EQ}{\begin{eqnarray}}
\newcommand{\EN}{\end{eqnarray}}
\newcommand{\EQQ}{\begin{eqnarray*}}
\newcommand{\ENN}{\end{eqnarray*}}
\newcommand{\R}{\mathbb R}

\renewcommand{\t}{^{\mbox{\tiny\sf T}}}

\newcommand{\bremark}
{\medskip\begin{remark}
\begin{rm}}
\newcommand{\eremark}{ \end{rm}\hfill \rule{1mm}{2mm}
\end{remark} }
\newcommand{\btheorem}{\medskip\begin{theorem} \begin{it}}
\newcommand{\etheorem}{\end{it} \hfill \rule{1mm}{2mm}
\end{theorem} }
\newcommand{\blemma}{\medskip\begin{lemma} \begin{it} }
\newcommand{\elemma}{ \end{it} \hfill\rule{1mm}{2mm}
\end{lemma} }
\newcommand{\bcorollary}{\medskip\begin{corollary} \begin{it} }
\newcommand{\ecorollary}{ \end{it} \hfill\rule{1mm}{2mm}
\end{corollary} }
\newcommand{\bdefinition}{\medskip\begin{definition} }
\newcommand{\edefinition}{ \hfill\rule{1mm}{2mm}
\end{definition} }
\newcommand{\bproposition}{\medskip\begin{proposition} }
\newcommand{\eproposition}{\hfill \rule{1mm}{2mm}
\end{proposition} }
\newcommand{\bexample}{\medskip\begin{example} \begin{rm}}
\newcommand{\eexample}{ \end{rm} \hfill\rule{1mm}{2mm}
\end{example} }

\newcommand{\basm}{\medskip\begin{assumption} \begin{rm} }
\newcommand{\easm}{ \end{rm} \hfill\rule{1mm}{2mm} \medskip
\end{assumption} }

\newtheorem{theorem}{\sf\bfseries Theorem}[section]
\newtheorem{lemma}{\sf\bfseries Lemma}[section]

\newtheorem{definition}{\sf\bfseries Definition}[section]
\newtheorem{remark}{\sf\bfseries Remark}[section]
\newtheorem{corollary}{\sf\bfseries Corollary}[section]
\newtheorem{proposition}{\sf\bfseries Proposition}[section]
\newtheorem{example}{\sf\bfseries Example}[section]
\newtheorem{assumption}{\sf\bfseries Assumption}

\title{\LARGE A Distributed Adaptive Scheme for
Multi-Agent Systems
\thanks{The work was supported by the Australian
Research Council under grant No. DP150103745. }}

\author{Imil Hamda Imran, Zhiyong Chen, Lijun Zhu, and  Minyue Fu % <-this % stops a space
\thanks{Imran, Chen, and Fu are with the School of Electrical Engineering and Computing,
        The University of Newcastle, Callaghan, NSW 2308, Australia. Zhu is with the Department of Electrical and
Electronic Engineering,  University of Hong Kong, China.
        {\tt\small
       ImilHamda.Imran@uon.edu.au,
        zhiyong.chen@newcastle.edu.au, ljzhu@eee.hku.hk,
        minyue.fu@newcastle.edu.au.}  }%
}

\begin{document}

\maketitle

 \begin{abstract}
 In traditional adaptive control, the certainty equivalence principle suggests a two-step design scheme.
A controller is first designed for the ideal
 situation assuming the uncertain parameter was known and it renders a Lyapunov function.
 Then, the uncertain parameter in the controller is replaced by its estimation that is updated by
 an adaptive law along the gradient of Lyapunov function.
 This principle does not generally work for a multi-agent system as an adaptive law based on the
 gradient of (centrally constructed) Lyapunov function cannot be implemented in a distributed fashion, except
 for limited situations.
In this paper, we propose a novel distributed adaptive scheme, not
relying on gradient of Lyapunov function, for general multi-agent
systems. In this scheme, asymptotic consensus of a
second-order uncertain multi-agent system is achieved in a network of directed graph.

\end{abstract}


\section{Introduction}

Control of multi-agent systems (MASs) is motivated by collective phenomena in natural systems
and extensive engineering applications, for example, cooperative
control of multiple unmanned aerial vehicles (UAVs) and mobile robots, distributed sensor networks, load balancing, etc.
Consensus is one of the most active research topics in
MASs from the systems and control perspective and it has achieved rapid progress in recent
years \cite{knorn2016}. The goal is to design collective algorithms for a group of agents such that
they achieve agreement in a certain sense of obeying common
dynamics.  A distributed consensus control protocol can generate effective
local control for each agent based on the relative measurement from
its neighbors via a network.

The research for MASs of homogeneous linear dynamics is mature
with the early works traced back to those on single or double
integrators. For example,
an observer based output feedback consensus  controller was constructed in
\cite{Scardovi2009} with both agent outputs and observer states
transmitted via network. 
  For a lower cost network with only output transmitted, consensus
protocols were studied in \cite{Tuna2009,Ma2010}.
 A low-gain approach can also be found  in
\cite{Seo2009} using a stable dynamic filter.
More general formation control for linear dynamics can be found in, 
e.g., \cite{Lin2016b, Lin2016}.

In many practical situations, agent dynamics are usually subject to uncertainties
that also induce heterogeneity.  To handle system uncertainties, an internal model based
approach has been proved to be effective. For example,
linear internal model based consensus techniques can found in
\cite{Wieland2011,Kim2011,Lunze2012,Grip2013} in different settings.
The basis idea is to introduce a reference trajectory for each agent
and collectively synchronize these references and hence agent outputs.

While certain nonlinearities of agent dynamics might be handled by
feedforward compensation, see, e.g., \cite{chen14b},  uncertain
nonlinearities likely bring more technical challenges. Most
existing results are also based on internal model design. For
instance, in \cite{Zhu2014}, the authors designed controllers for
MASs of second-order nonlinear dynamics with agreement on a
constant. More general nonlinear dynamics were studied in
\cite{Isidori2014,Chen:2014} that require that all agents exchange
full state information. The most sophisticated  result was given
in \cite{zhu2016} in the output communication setting using a
small gain theorem. Some other relevant internal model design can
be found for cooperative output regulation in a leader-following
setting; see, e.g. \cite{Su2013a,Huang2015}.

Another research line is to deal with system uncertainties, in particular, unknown
parameters, using adaptive control.
Like in traditional adaptive control, the certainty equivalence principle suggests a two-step design scheme.
A controller is first designed for the ideal
 situation assuming the uncertain parameter was known and it renders a Lyapunov function.
 Then, the uncertain parameter in the controller is replaced by its estimation that is updated by
 an adaptive law along the gradient of Lyapunov function.

In literature, such an adaptive control scheme has been
investigated for MASs in some situations. For example, a
first-order MAS was studied in \cite{wang2012} for a network of
undirected graph. The result was presented in a more general
framework in \cite{chen2014a}. Similar adaptive technique was used
in \cite{chen2014b} for both first-order and second-order MASs
with a Nussbaum gain added to deal with unknown control direction.
Also for networks of undirected graphs, but under the jointly
connected condition, an adaptive scheme was studied for
first-order MASs in \cite{yu2012} and \cite{yu2017} for
leader-following and leaderless settings, respectively. In
particular,  in \cite{yu2012} each agent requires ``not only the
information of its neighbors but also the information of its
neighbors' neighbors'' and then in \cite{yu2017}  the approach was
improved to a purely distributed design. 
 It is noted that adaptive control was also used to 
tune the coupling weights of a network in, e.g.,  \cite{li2013}.

 For a network of directed graph, the associated Laplacian is asymmetric, which significantly complicates the problem.
 Some relevant work can be found in
  \cite{zhang2012} that gave a result for higher-order MASs, but for the leader-following case.
Moreover,  it is noted that consensus in \cite{zhang2012} cannot be achieved asymptotically but with a residual error.
 The work in  \cite{zhang2012} also considers neural network (NN) approximation for the unknown nonlinearities.
 The  residual error  is caused not only by NN approximation error,
 but also by the cost of distributed implementation of the adaptive law.
 In other words, the residual consensus error still exists even if the NN error is zero.
 The work in  \cite{zhang2012} includes the early results in \cite{das2010, das2011} as special cases.

Even though an adaptive law along the gradient of Lyapunov
function using the certainty equivalence principle  has been
proved to be successful in the aforementioned scenarios, it does
not work for MASs in general as a Lyapunov function is usually
centrally constructed. In other words, distributed implementation
of the gradient of Lyapunov function is usually impractical except
fort limited cases.  For instance,  it still remains open to
design a distributed adaptive law to achieve asymptotic consensus
for a second-order MAS in a directed network.  As will be
explained in detail later in this paper, an adaptive law along the
gradient of Lyapunov function has its inherent drawback to solve
this open problem due to the lack of its distributed
implementation.

In this paper, we propose a novel distributed adaptive scheme, not
relying on gradient of Lyapunov function, for general MASs. In the
gradient based scheme, the estimation error is expected to have a
steady state zero.  To drive the agent states together with the
estimation error to their steady states, the adaptive law must
follow the gradient of Lyapunov function. The novel idea is to
introduce an input compensation such that the steady state of the
estimation error is not zero but a manifold in the state space of
agent states and estimated parameters. By proper selection of the
manifold, it  can be made attractive without relying the centrally
designed Lyapunov function. At the manifold, the agent states also
approach their desired steady state. The idea in characterizing
the steady-state manifold originates from the steady-state
generator design in the output regulation theory for dealing with
asymptotic disturbance rejection and reference tracking
\cite{huang2004,Chen2015Book} and immersion and invariance
adaptive control of nonlinearly parameterized systems
\cite{liu2010}. Within the novel distributed adaptive scheme, the
aforementioned open problem on asymptotic consensus of a
second-order nonlinear MAS in a directed network is solved.

\section{Preliminaries and Motivating Examples}

Consider a network of MAS with a properly designed controller, described
by
\EQ
\dot x_i=f_i(x), \; i=1,\cdots, n,\label{cl0}\EN
where $x_i \in \mathbb{R}^{l}$ is the state of the $i$-th agent and 
$f_i(x)$ is a general function representing the agent dynamics. 
Denote $x=[x_1\t, x_2\t,\cdots,x_n\t ]\t$ and $f(x)=[f\t_1(x), f\t_2(x),\cdots,f_n\t(x) ]\t$. So, the network
has the compact form $\dot x = f(x)$. This is the nominal closed-loop MAS free of uncertainties.
Suppose the MAS has achieved a certain consensus behavior, specifically, with a property
in term of a Lyapunov-like function. 
 Throughout the note, the notation $\| \cdot \|$ means the Euclidean norm and 
$\|x\|_R =\| R x\|$ for a real matrix $R$.

\basm \label{ass-LP} There exists a continuously differentiable function $V(x)$
satisfying $ \underline\alpha(\|x\|_R)  \leq V(x)  \leq \bar\alpha(\|x\|_R)$  
for a matrix $R\in{\mathbb R}^{\bar nl\times nl}$  with $\bar n\leq n$ and class ${\cal K}_\infty$ functions
$\underline\alpha$ and $\bar\alpha$, such that,
\EQ
 \frac{\partial V(x)}{\partial x} f(x) \leq -\alpha(\|x\|_R)
\EN
for a class ${\cal K}_\infty$ function $\alpha$.  Moreover,  
\EQ
 \frac{  \left\|\frac{\partial V(x)}{\partial x} \right\|^2}{\alpha(\|x\|_R) } \leq \sigma 
\EN for some positive constant $\sigma$.
\easm

\bremark Two typical scenarios of Assumption~\ref{ass-LP}  are explained as follows. 

(i) If $R\in{\mathbb R}^{nl\times nl}$, i.e., $\bar n =n$, is a nonsingular matrix, 
then $\|x\|_R =0$ implies $\|x\|=0$. In this scenario, the function $V(x)$ is a Lyapunov function
for the $x$-system and Assumption~\ref{ass-LP} implies $\lim_{t\rightarrow\infty} \|x(t)\| =0$, i.e., 
asymptotic stability about the equilibrium at the origin.  

(ii) If $R \in \R^{(n-1)l\times nl}$, i.e., $\bar n=n-1$, is a full row rank matrix and the rows are perpendicular to
${\rm span} \{ {\bf 1} \otimes I_l \}$ where $I_l \in {\mathbb R}^{l\times l}$ is an identity matrix and 
${\bf 1}=\left[\begin{array}{ccc}
1 & \ldots & 1\end{array}\right]\t \in\mathbb{R}^{n}$, then
$\|x\|_R =0$ implies $x=  {\bf 1} \otimes x_o$ for some $x_o\in\R^l$.
In this scenario, the function $V(x)$ is a Lyapunov function
for the $Rx$-subsystem and Assumption~\ref{ass-LP} implies $\lim_{t\rightarrow\infty} \|x(t)\|_R =0$, i.e., 
 $\lim_{t\rightarrow\infty} [x(t) - {\bf 1} \otimes x_o(t)] =0$, which is a typical  consensus phenomenon.

\eremark

Now, we consider the network subject to uncertainties and the objective is
to design an adaptive scheme to deal with the uncertainties such that the behavior of the nominal
system is still maintained.
The design of an adaptive law is expected to be separated from the consensus controller in
the nominal system,  which is not explicitly shown in the closed-loop structure (\ref{cl0}).

Specifically,  the network of MAS subject to uncertainties is represented by
 \EQ
\dot x_i=f_i(x)+ g_i(x_i,w_i ,\mu_i ), \; i=1,\cdots, n \label{plant-adp}\EN
where $w_i \in \mathbb{R}^{m_i}$ represents constant unknown parameters and  $\mu_i\in \mathbb{R}^{m_i}$  an additional
control input to adaptively account for the uncertainties.
Suppose the uncertainties have the linearly parameterized structure, i.e.,
\EQ
g_i(x_i, w_i,\mu_i)=  h_i(x_i) (w_i - \mu_i). \label{lp}\EN
 for some function $h_i(x_i)$.
We can rewrite the system in a compact form
\EQ
\dot x=f(x)+ H(x) (w - \mu) \EN
where
$w=[w_1\t, w_2\t,\cdots,w_n\t ]\t$,
 $\mu=[\mu_1\t, \mu_2\t,\cdots,\mu_n\t ]\t$, and
 $H(x)=\mbox{diag}\left[\begin{array}{cccc}
h_1(x_1) & h_2(x_2)  & \cdots & h_n(x_n) \end{array}\right].$

If the parameter $w_i$ were known,  $\mu_i = w_i$ could trivially
cancel the uncertainties $g_i(x_i,w_i,\mu_i)$. For the practical case
with an unknown $w_i$, an adaptive law can be designed along the gradient
of the Lyapunov function $V(x)$, as summarized as follows.

\btheorem \label{thm-cen} {\bf (Centralized Scheme)}
For the system (\ref{plant-adp}) with (\ref{lp}) under Assumption~\ref{ass-LP},
with the controller
\EQ
\mu &=& \hat w \nonumber\\
\dot{\hat w}\t &=&  \lambda \frac{\partial V(x)}{\partial x} H(x)  ,\; \lambda>0 \label{adp}
\EN
the derivative of 
\EQQ U(x, \tilde w) =V(x) +\tilde w\t \tilde w / (2\lambda) \ENN
 with $\tilde w =\hat w -w$ satisfies
\EQ
\dot U(x, \tilde w) \leq  -\alpha(\|x\|_R),
\EN
along the trajectory of the closed-loop system
 (\ref{plant-adp})+(\ref{lp})+(\ref{adp}).
\etheorem

{\it Proof:} Direct calculation shows that the derivative of
$V(x)$ along the dynamics (\ref{plant-adp}) with (\ref{lp}) satisfies
\EQQ
\dot V(x) %= \frac{\partial V(x)}{\partial x} f(x) + \frac{\partial V(x)}{\partial x} h(x)(w-\hat w)\\
= \frac{\partial V(x)}{\partial x} f(x) - \frac{\partial V(x)}{\partial x} H(x)\tilde w \\
\leq -\alpha(\|x\|_R) - \frac{\partial V(x)}{\partial x} H(x)\tilde w . \ENN
Hence,
\EQQ
\dot U(x,\tilde w)  \leq -\alpha(\|x\|_R) - \frac{\partial V(x)}{\partial x} H(x)\tilde w  +  \dot{ \tilde w}\t \tilde w /\lambda \\
= -\alpha(\|x\|_R)  \ENN
for $\dot{\tilde w} =\dot {\hat w}$ given in (\ref{adp}).
\eproof

The adaptive law (\ref{adp}) can be rewritten as follows, for $i=1,\cdots, n$,
\EQ
\mu_i &=& \hat w_i \nonumber\\
\dot{\hat w}_i\t &=&  \lambda \frac{\partial V(x)}{\partial x_i} h_i(x_i),\; \lambda>0 \label{adpi}
\EN
that is not always distributed as ${\partial V(x)}/{\partial x_i}$ depends on not only the local state
of agent $i$,
but also the full network state $x$ unless $V(x)$ can be properly designed to have
a distributed ${\partial V(x)}/{\partial x_i}$
on a case by case basis.  However, it can be true only for very limited cases because
the function $V(x)$ for the nominal system is constructed in a centralized manner.
Two motivating examples are given as follows.

\bexample
Consider a first-order integrator MAS in a network of an {\it undirected} graph associated with a symmetric Laplacian $L$.
The nominal network dynamics are $\dot x = -Lx$ for $x\in{\mathbb R}^{n}$. Let $V(x)=x\t Lx/2$ where
 $L= R\t R$ for a full row rank matrix $R\in{\mathbb R}^{(n-1)\times n}$ when the
graph is connected.  The derivative along the trajectory of $\dot x = -Lx$ is
$\dot V(x) =- x\t L L x  =- (Rx)\t (R R\t) (R x)  \leq - r_{\rm min} \|x\|^2_R$ where $r_{\rm min}>0$ is the minimal eigenvalue of $R R\t$. When the network is subject to uncertainties $H(x)w$, i.e.,
$\dot x = -Lx +H(x) (w-\mu)$, following Theorem~\ref{thm-cen},
the additional adaptive controller $\mu$ in (\ref{adp}) has the specific form
\EQQ
\dot{\hat w}_i &=&  \lambda  h\t_i(x_i)  L_i x \\
&=& \lambda h_i\t (x_i) \sum_{j\in {\cal N}_i \cup \{i\}}
 l_{ij} x_j  ,\; \lambda>0,\ENN
with $L_i$ the $i$-th row of $L$ and ${\cal N}_i$ the set of neighbors of $i$. In this scenario,
the adaptive scheme is implemented in a distributed fashion. This development can be found in, e.g., \cite{yu2017}.
\eexample

\bexample
Consider a first-order integrator MAS
$\dot x = -Lx$ in a network of a {\it directed} graph
associated with a Laplacian of the special form
\EQ\label{leaderL}
L= \left[ \begin{array}{cc}
0 & 0_{1\times (n-1)}\\
-b & L_o+B
 \end{array}\right]
\EN
that represents a {\it leader-following} network with agent $1$ as the leader.
The matrix $L_o$ is the Laplacian of the sub-network of followers  and $B=\mbox{diag}(b), b=[b_2,\cdots, b_n]\t$
with $b_i \geq 0$ the  weight from the leader to agent $i$. Denote $L_o=D-E$ where $D=\mbox{diag}(d_2,\cdots, d_n)$ a
diagonal matrix and $E$ an off-diagonal one.  Assume the network has a spanning tree with
the root node being the leader node $1$. Then, there exists a diagonal matrix $P=\mbox{diag}(p_2,\cdots,p_n)>0$ such that
\EQQ 2 Q= P(L_o+B) +(L_o+B)\t P>0.\ENN
Let \EQQ
R= \left[ \begin{array}{cc}
-b & L_o+B
 \end{array}\right],
\ENN
one has
 \EQQ
RL = (L_o+B)R.
 \ENN
Let $V(x)= x\t R\t  P Rx/2$.  The derivative along the trajectory of $\dot x = -Lx$ is
\EQQ
\dot V(x)=-x\t R\t  [P (L_o+B) +(L_o+B)\t P ]R x/2 \leq -x\t R\t Q R x.
\ENN
 When the network is subject to uncertainties $H(x)w$, i.e.,
$\dot x = -Lx + H(x) (w-\mu)$, along which
the derivative of $U(x, \tilde w) =V(x) +\tilde w\t \tilde w / (2\lambda)$ is \EQQ
\dot U(x) \leq -x\t R\t Q R x
+ [\dot {\hat w} \t  /\lambda-  x\t R\t  P R  H(x)] \tilde w.
\ENN
Following Theorem~\ref{thm-cen},
the update law in (\ref{adp}) has the specific form
\EQQ
\dot{\hat w} =  \lambda   H\t(x) R\t P R x
\ENN
that however cannot be implemented  in distributed fashion. In fact, a distributed adaptive
law for this scenario still remains open.

For the scenario studied in \cite{zhang2012},  the leader is free of uncertainty, i.e., $h_1(x) =0$ and
$w\in\R^0$ trivially. Then,
one has
\EQQ
R H(x)= \left[ \begin{array}{cc}
-b & L_o+B
 \end{array}\right]  \mbox{diag}\left[\begin{array}{cccc}
0 & h_2(x_2)  & \cdots & h_n(x_n) \end{array}\right] \\
=
(L_o+B) \bar H(x) = (D+B) \bar H(x) -E \bar H(x)
\ENN
for $\bar H(x)=\left[0, \mbox{diag}\left[\begin{array}{cccc}
h_2(x_2)  & \cdots & h_n(x_n) \end{array}\right]\right].$
The following update law was applied
\EQQ
 \dot {\hat w} =   \lambda \bar H\t(x) (D+B) P Rx -\lambda \kappa \hat w
\ENN 
that gives
\EQQ
\dot U(x) \leq -x\t R\t Q R x
+ [-\kappa \hat w\t + x\t R\t  P E \bar H(x)] \tilde w \\
\leq \left[ -x\t R\t Q R x
  + x\t R\t  P E \bar H(x)  \tilde w -\kappa \|\tilde w\|^2 \right] + \kappa \|w\|  \|\tilde w\|.
\ENN
The update law is implemented in a distributed fashion by noting that the matrices $P, D$
and $B$ are diagonal, that is, 
\EQQ
 \dot {\hat w}_i =  \lambda (d_i+b_i) p_i h_i\t(x)  L_i x -\lambda\kappa \hat w_i, \;, i=2,\cdots, n,
\ENN
with $L_i$ the $i$-th row of $L$.
In the expression of $\dot U(x)$, the terms in the square brackets
can be made negative with a sufficiently large $\kappa$ but the positive term $\kappa \|w\|  \|\tilde w\|$
causes a residual consensus error. In other words, no asymptotic consensus can be achieved using
the approach developed in \cite{zhang2012}.
\eexample

\section{A Distributed Adaptive Scheme}

The main contribution of this paper is to bring a novel adaptive scheme that can be implemented
in a distributed fashion. For this purpose, let us have a close inspection on the approach in
Theorem~\ref{thm-cen}.
For the system (\ref{plant-adp}) with linearly parameterized uncertainties, we introduce a virtual exosystem
\EQ
\dot \tau _i &=&f_i(\tau)  \nonumber \\
\dot w_i &=&0, \;i=1,\cdots,n.\label{exo}
\EN
The agent state $x_i$ and input $\mu_i$ are expected to have the steady states $x_{i,ss}=\tau_i$ and $\mu_{i,ss}=w_i$, respectively.
In this sense, we call
\EQQ
\dot w_i &=&0, \\
\mu_{i,ss}&=&w_i, \;i=1,\cdots,n
\ENN
the steady-state generator for the input $\mu_i$, which motivates the update law
 \EQQ
\dot {\hat w}_i &=&0 + \nabla, \\
\mu_i &=&\hat {w}_i, \;i=1,\cdots,n
\ENN
where $\nabla$ is designed along the gradient of Lyapunov function such that the manifold
$\{ (x,\mu,\tau,w)\;  | \; x_i =\tau_i,\; \mu_i =w_i, \; i=1,\cdots, n \}$ is attractive.

The novel idea is to introduce a function $\beta_i(x_i)$ to the input, i.e.,
$\mu_i =-\beta_i(x_i) +\hat \mu_i$.
Along the virtual exosystem (\ref{exo}),
the agent state $x_i$ and input $\hat\mu_i$ are expected to have the steady states
$x_{i,ss}=\tau_i$ and $\hat\mu_{i,ss}= \theta_i(\tau_i,w_i) =\beta_i (\tau_i)+w_i$, respectively.
As a result, we have a steady-state generator for the input $\hat\mu_i$
 \EQQ
\dot \theta_i(\tau_i,w_i) &=& \frac{\partial \beta_i(\tau_i)}{\partial \tau_i} f_i(\tau) \\
{\hat \mu}_{i,ss} &=& \theta_i(\tau_i,w_i) , \;i=1,\cdots,n,
\ENN
that motivates the update law
 \EQQ
\dot {\hat w}_i &=& \frac{\partial \beta_i(x_i)}{\partial x_i} f_i(x) \\
{\hat \mu}_{i} &=& \hat w_i,\; \;i=1,\cdots,n.
\ENN
In this design,  $\beta_i$ can be properly selected such that the manifold
$\{ (x,\hat\mu,\tau,w)\;  | \; x_i =\tau_i, \; \hat\mu_i = \theta_i(\tau_i,w_i),\;i=1,\cdots,n \}$ is attractive.
The introduction of  $\beta_i$ avoids the implementation of $\nabla$ that relies on a centrally designed
Lyapunov function.

In this new development, if we treat $\hat w_i$ as the estimated value of $w_i$, the steady state
of the estimation error $\hat w_i-w_i$ is  not zero but $\theta_i(\tau_i,w_i) -w_i =\beta_i(\tau_i)$
where $\tau_i$ is the steady state of $x_i$.
Therefore, we aim to drive $\hat w_i-w_i$ to the  manifold
$\{ (x_i, \hat w_i) \; |\; \hat w_i-w_i =\beta_i(x_i), \; i=1,\cdots, n \}$
in the space of agent states and estimated parameters.
By proper selection of the manifold, it can be made attractive and
the agent state $x_i$ can approach its desired steady state $\tau_i$ on the manifold.
The rigorous formulation of the approach is given in the following theorem.

 \btheorem {\bf (Distributed Scheme)} \label{thm-dis}
 Consider the system (\ref{plant-adp}) with (\ref{lp})
under Assumption~\ref{ass-LP}. 
 Let the distributed controller be
\EQ
\mu_i &=& \hat w_i-\beta_i(x_i) \nonumber\\
\dot{\hat w}_i &=& -\lambda_i h\t_i(x_i)  f_i(x)   \label{adp-ds}
\EN
where $\beta_i(x_i)$ is any continuously differentiable function satisfying
\EQ
\frac{\partial \beta_i(x_i) }{\partial x_i} =-\lambda_i h\t_i(x_i), \label{partialbetai}
\EN
for some $\lambda_i >0$.
Then, the derivative of
\EQQ U(x, z) =V(x) + \frac{\sigma}{4(1-k)}\sum_{i=1}^n z_i\t z_i /(2\lambda_i),
\ENN with
\EQ
z_i =  \beta_i(x_i) -\tilde w_i,  \; \tilde w_i =\hat w_i -w_i,
\EN satisfies
\EQ
\dot U(x, z) \leq  -k \alpha(\|x\|_R),
\EN
for any $0<k<1$, along the trajectory of the closed-loop system
 (\ref{plant-adp})+(\ref{lp})+(\ref{adp-ds}). 
% Moreover, $\lim_{t\rightarrow\infty} h_i(x_i(t)) z_i(t) =0$.

 \etheorem

{\it Proof:} The system composed of  (\ref{plant-adp})+(\ref{lp})+(\ref{adp-ds})
can be rewritten as
 \EQQ
\dot x_i=f_i(x)+ h_i(x_i)(w_i -\hat w_i +\beta_i(x_i) ) \\
=f_i(x)+ h_i(x_i)z_i. \ENN
Direct calculation shows
\EQQ
\dot V(x) = \frac{\partial V(x)}{\partial x} f(x) + \sum_{i=1}^n \frac{\partial V(x)}{\partial x_i} h_i(x_i)z_i\\
\leq -\alpha(\|x\|_R)   + \sum_{i=1}^n \frac{\partial V(x)}{\partial x_i} h_i(x_i)z_i.
\ENN
For any $0<k<1$, pick  $a = (1-k)/{\sigma}$. One has
 \EQQ
   a \left\|\frac{\partial V(x)}{\partial x} \right\|^2
\leq (1-k) \alpha(\|x\|_R).
\ENN
Moreover
\EQQ
\dot V(x)  \leq -\alpha(\|x\|_R)   + \sum_{i=1}^n \left\{ a \left\|\frac{\partial V(x)}{\partial x_i} \right\|^2  + \|h_i(x_i)z_i\|^2 /(4a) \right\}\\
\leq -\alpha(\|x\|_R)   + a \left\|\frac{\partial V(x)}{\partial x} \right\|^2  +  \sum_{i=1}^n  \|h_i(x_i)z_i\|^2 /(4a) \\
\leq -k \alpha(\|x\|_R)      +  \sum_{i=1}^n  \|h_i(x_i)z_i\|^2 /(4a) .
\ENN

Next, one has
\EQQ
\dot z_i &=&
\frac{\partial \beta_i(x_i) }{\partial x_i} \dot x_i -\dot {\hat w}_i \\
&=& \frac{\partial \beta_i(x_i) }{\partial x_i} f_i(x) +\frac{\partial \beta_i(x_i) }{\partial x_i}
h_i(x_i)z_i -  \frac{\partial \beta_i(x_i) }{\partial x_i} f_i(x)  \\
&=& \frac{\partial \beta_i(x_i) }{\partial x_i}
h_i(x_i)z_i  =- \lambda_i h\t_i(x_i)  h_i(x_i)z_i.
 \ENN
 Then, the derivative of  $z_i\t z_i/(2\lambda_i)$ along the above trajectory
 is
 \EQQ
 \frac{d(z_i\t z_i/ (2\lambda_i) )}{dt} = -z_i\t h\t_i(x_i)  h_i(x_i)z_i =-\|h_i(x_i)z_i\|^2.
 \ENN
As a result, the derivative  of
 \EQQ U(x, z) =V(x) + \frac{1}{4a}\sum_{i=1}^n z_i\t z_i /(2\lambda_i),
\ENN  along the trajectory of the closed-loop system, is
 \EQQ
 \dot U(x,z) &\leq& -k \alpha(\|x\|_R)      +  \sum_{i=1}^n  \|h_i(x_i)z_i\|^2 /(4a) \\
&& -   \sum_{i=1}^n  \|h_i(x_i)z_i\|^2 /(4a) \\
&\leq&  -k \alpha(\|x\|_R).
 \ENN
 The proof is thus completed. \eproof

\bremark
In Theorem~\ref{thm-dis} the adaptive controller (\ref{adp-ds}) is implemented at each agent $i$.
This scheme is distributed as it only relies on the agent state $x_i$ and its nominal dynamics $f_i(x)$.
The nominal dynamics $f_i(x)$ is implemented before hand  for the
ideal situation free of uncertainties, typically in distributed fashion. The effectiveness of Theorem~\ref{thm-dis} 
will be demonstrated by a network of second-order uncertain dynamics in the next section.
\eremark

\section{Network of Second-Order Uncertain Dynamics}

We consider a group of $n\geq 2$ agents governed by a set of second-order nonlinear
differential equations
\EQ
\dot{p}_{i} & = & v_{i}\nonumber \\
\dot{v}_{i} & = & \alpha_{1}p_{i}+\alpha_{2}v_{i}+\xi_{i}\left(w_{i},v_{i}\right)+u_{i}, \;  i=1,\ldots,n, \label{agents}
\EN
where $p_{i},v_{i} \in \R$ are the states and $u_i\in\R$ is the input of the agent $i$.
The function $\xi_{i}(w_{i},v_{i}) = \zeta_i (v_i) w_i$ for a bounded function $\zeta_i(v_i)$
represents heterogeneous nonlinearities with
$w_{i}$ an unknown constant parameter.
The two parameters $\alpha_1$ and  $\alpha_2$ are known.
For convenience of presentation, we denote
 $$ A=\left[\begin{array}{cc}
0 & 1\\
\alpha_{1} & \alpha_{2}
\end{array}\right],\;\; x_{i}=\left[\begin{array}{c}
p_{i} \\ v_{i}\end{array}\right]$$
and
$$
p=\left[\begin{array}{c}
p_{1} \\ \vdots \\ p_n
\end{array}\right],
v=\left[\begin{array}{c}
v_{1} \\ \vdots \\ v_n
\end{array}\right],
x=\left[\begin{array}{c}
x_{1} \\ \vdots \\ x_n
\end{array}\right],
u=\left[\begin{array}{c}
u_{1} \\ \vdots \\ u_n
\end{array}\right].
$$

In this section, the network topology is given by a directed graph
$\mathcal{G}=\{\mathcal{V},\mathcal{E}\}$, where $\mathcal{V}=\{1,\cdots,n\}$ denotes a finite non-empty set of nodes (i.e., agents) and $\mathcal{E}\subset\mathcal{V}\times\mathcal{V}$ presents the set of edges (i.e., communication links).
The adjacency matrix ${\cal A}=[a_{ij}]$ of a weighted directed graph is defined as $a_{ii}=0$ (no self-loop) and $a_{ij}>0$ if $(j,i)\in\mathcal{E}$ where $i\neq j$. Let the Laplacian $L$ be defined as $L_{ii}=\sum_{j\neq i}a_{ij}$ and $L_{ij}=-a_{ij}$, where $i\neq j$.
For a distributed algorithm, each agent $i$ can achieve the information from the network as follows,
with $L_i$ the $i$-th row of $L$,
\EQQ
L_i p= - \sum_{j=1}^n a_{ij} (p_j -p_i)\\
L_i v= - \sum_{j=1}^n a_{ij} (v_j -v_i).
\ENN
In this section, we study a general directed leaderless setting that includes the leader-following case (with the
Laplacian of the special form (\ref{leaderL})) as a special case.  Throughout the section, we have the following assumption.

\basm \label{ass-spanningtree}
The network has a directed spanning tree.
\easm

The objective is to design a distributed adaptive consensus protocol (i.e., only $p_i$, $v_i$,
$L_i p$ and $L_i v$ are available measurements for agent $i$) under Assumption~\ref{ass-spanningtree}, 
such that the MAS has the following asymptotic property
\EQ  \lim_{t\rightarrow \infty} p(t)- p_o(t){\bf 1} =0 \nonumber\\
\lim_{t\rightarrow \infty} v(t)- v_o(t){\bf 1} =0 \label{consensus}
\EN
for some time functions $p_o(t), v_o(t): [0,\infty)\mapsto \R$.

Under Assumption~\ref{ass-spanningtree}, the Laplacian $L$
has one zero eigenvalue and the other eigenvalues have positive
real parts. Let the vectors $r\in\mathbb{R}^{n}$ and ${\bf 1}$ be
the left and right eigenvectors corresponding to the eigenvalue
zero of $L$, in particular, $r\t L=0$, $L  \mathbf{1} =0$, and $r\t
{\bf 1}=1$.

 There exist matrices $W\in\mathbb{R}^{\left(n-1\right)\times n}$,
$U\in\mathbb{R}^{n \times \left(n-1\right)}$ such that
 \EQQ
T =  \left[ \begin{array}{c} r\t \\ W \end{array} \right] ,\;
T^{-1} =  \left[\begin{array}{cc}
{\bf 1}& U \end{array}\right].
\ENN
One has the following similarity transformation
\EQQ
TLT^{-1}=\left[\begin{array}{cc}
0 & 0\\
0 & J
\end{array}\right]
\ENN
where $J=WLU\in\mathbb{R}^{\left(n-1\right)\times\left(n-1\right)}$
is a matrix with all eigenvalues having positive real parts.
Define the matrix $R$ as follows
\EQQ \left[\begin{array}{c}
Wp\\
Wv
\end{array}\right] =R x.
\ENN
It is easy to check that $R$ has a full row rank and  the rows of $R$
are perpendicular to
${\rm span} \{ {\bf 1} \otimes I_2 \}$.

 We have the following technical lemma that has been used in \cite{hou2017}
with the proof hidden in system analysis. A direct proof on matrix analysis is given in 
appendix for readers' convenience. 

\blemma \label{lemmaA} Under Assumption~\ref{ass-spanningtree},
there exist $\gamma_1, \gamma_2>0$ such that the matrix
$\bar{A}=\left[\begin{array}{cc}
0 & I\\
\alpha_{1}I-\gamma_{1}J & \alpha_{2}I-\gamma_{2}J
\end{array}\right]$ is Hurwitz.
\elemma

\medskip

The next lemma shows the consensus result for the ideal situation. 

\blemma Under Assumption~\ref{ass-spanningtree}, consider the system (\ref{agents})  with $\xi_i(w_i,v_i)=0$ and
\EQ
u_i =  -\gamma_{1}L_i p-\gamma_{2}L_i v, \label{ui-ideal}
\EN
where $\gamma_1$ and $\gamma_2$ are such that
the matrix $\bar{A}=\left[\begin{array}{cc}
0 & I\\
\alpha_{1}I-\gamma_{1}J & \alpha_{2}I-\gamma_{2}J
\end{array}\right]$ is Hurwitz. Let
$P = P\t >0$ be the solution to the Lyapunov equation
\EQQ P\bar{A}+\bar{A}\t P =-I.
\ENN
The function
\EQ V(x)= x\t R\t P R x
\EN
satisfies $P_{\rm min}\|x\|^2_R \leq  V(x) \leq P_{\rm max}  \|x\|^2_R$
($P_{\rm min}$ and $P_{\rm max}$ are the minimum and maximum eigenvalues of $P$)
and its derivative
along the closed-loop system is
\EQ \dot V(x) =- \|x\|^2_R.
\EN
\label{lemma-ideal}
\elemma

{\it Proof:}
The closed-loop system composed of (\ref{agents})  and (\ref{ui-ideal}) is
\EQ
\dot{p}_{i} & = & v_{i}\nonumber \\
\dot{v}_{i} & = & \alpha_{1}p_{i}+\alpha_{2}v_{i} -\gamma_{1}L_i p-\gamma_{2}L_i v , \;  i=1,\ldots,n, \label{xi-ideal} \EN
denoted as $\dot x_i =f_i(x).$ It can also be put  in a compact form
\EQ
\dot{p} & = & v \nonumber \\
\dot{v} & = & \alpha_{1}p +\alpha_{2}v  -\gamma_{1}L p-\gamma_{2}L v. \EN

From the definition of $T$ and $T^{-1}$, one has
\EQQ {\bf 1} r\t + U W =I
\ENN
and
\EQQ
J W =WL UW =WL (I-{\bf 1} r\t) =WL.
\ENN
Using this fact, we have the following calculation
\EQQ
 R\dot x
=    \left[\begin{array}{c}
W \dot p\\
W \dot v
\end{array}\right] \\
=    \left[\begin{array}{c}
W v \\
\alpha_{1}W p +\alpha_{2}W v  -\gamma_{1}W L p-\gamma_{2}W L v
\end{array}\right]\\
=    \left[\begin{array}{c}
W v \\
\alpha_{1}W p +\alpha_{2}W v  -\gamma_{1}J W p-\gamma_{2} JW v
\end{array}\right]\\
=   \left[\begin{array}{cc}
0 & I \\
\alpha_{1}I -\gamma_{1}J &  \alpha_{2}  I -\gamma_{2} J
\end{array}\right]  \left[\begin{array}{c}
W p\\
W v
\end{array}\right]
=    \bar A  Rx.
\ENN
 As a result,
 \EQQ
 \dot V(x) = x\t R\t P R\dot x + \dot x\t R\t P R x \\
= x\t R\t (P   \bar A +\bar A\t P)  Rx  =-\|x\|^2_R.
 \ENN
 The proof is completed.
  \eproof

The main result on a distributed adaptive controller is stated in the following theorem that 
is proved by applying Theorem~\ref{thm-dis}.

\btheorem \label{thm:mas} Under Assumption~\ref{ass-spanningtree}, consider the system (\ref{agents})  with the controller
\EQ
u_i =  -\gamma_{1}L_i p-\gamma_{2}L_i v - \zeta_i(v_i)\mu_i \label{controller-mu}
\EN
where $\gamma_1$ and $\gamma_2$ are given in Lemma~\ref{lemma-ideal},
\EQ
\mu_i &=& \hat w_i- \rho_i(v_i) \nonumber\\
\dot{\hat w} &=& -\lambda_i \zeta\t_i(v_i)  [\alpha_{1}p_{i}+\alpha_{2}v_{i} -\gamma_{1}L_i p-\gamma_{2}L_i v],  \label{adp-2nd}
\EN
and $\rho_i(v_i)$ is any continuously differentiable function satisfying
 \EQ
\frac{\partial \rho_i(v_i) }{\partial v_i} =-\lambda_i \zeta\t_i(v_i) , \; \lambda_i >0.  \label{partialrhoi}
\EN
Then, consensus is achieved in the sense of (\ref{consensus})
for some time functions $p_o(t), v_o(t): [0,\infty)\mapsto \R$.
\etheorem

{\it Proof:} The closed-loop system composed of (\ref{agents}) and  (\ref{controller-mu}) is, for  $i=1,\ldots,n$,
\EQ
\dot{p}_{i} & = & v_{i}\nonumber \\
\dot{v}_{i} & = & \alpha_{1}p_{i}+\alpha_{2}v_{i} -\gamma_{1}L_i p-\gamma_{2}L_i v +\zeta_i (v_i) (w_i-\mu_i), \;\label{pvi}
\EN
or in a compact form (\ref{plant-adp}), i.e., 
\EQQ
\dot x_i =f_i(x) +h_i(x_i) (w_i-\mu_i)
\ENN
where  $\dot x_i =f_i(x) $ is given in (\ref{xi-ideal}) and
\EQQ
h_i(x_i) =\left[ \begin{array}{c} 0 \\ \zeta_i (v_i) \end{array}\right].
\ENN

In Lemma~\ref{lemma-ideal}, it has been proved that Assumption~\ref{ass-LP} is satisfied for $\dot x_i =f_i(x)$.
It is noted that
\EQ
 \frac{  \left\|\frac{\partial V(x)}{\partial x} \right\|^2}{ \|x\|^2_R }
 = \frac{  \left\|2 x\t R\t P R  \right\|^2}{ \|x\|^2_R }
 \leq 4\|PR\|^2 < \infty.
\EN

For (\ref{partialrhoi}) and $\beta_i(x_i)= \rho_i(v_i)$, one has  (\ref{partialbetai}).
Also, (\ref{adp-ds}) takes the special form (\ref{adp-2nd}).
By Theorem~\ref{thm-dis}, one has
\EQ
\dot U(x, z) \leq  -k \|x\|^2_R
\EN
 for
\EQQ U(x, z) =V(x) + \frac{\sigma}{4(1-k)}\sum_{i=1}^n z_i\t z_i /(2\lambda_i),
\ENN and
$z_i =  \rho_i(v_i) -\tilde w_i,  \; \tilde w =\hat w -w$.

It is obvious to see that both
$\|x(t)\|_R$ and $z(t)$ are bounded.
Because of
\EQQ
R\dot x =\bar A R x +R H(x) z,
\ENN
$\|\dot x(t)\|_R$ is bounded and hence $-k \|x(t)\|^2_R$ uniformly continuous in $t$.
By Barbalat's Lemma, one has $\lim_{t\rightarrow \infty} \|x(t)\|_R =0$, that is,
\EQQ
\lim_{t\rightarrow \infty}  \left[\begin{array}{c}
W p(t)\\
W v(t)
\end{array}\right] =0 .
\ENN
Let $p_o(t)=r\t p(t)$ and $v_o(t)=r\t v(t)$.
From the following relationship
\EQQ
p  =  \left[\begin{array}{cc}
{\bf 1}& U\end{array}\right] \left[\begin{array}{c}
r\t p\\
Wp
\end{array}\right]= {\bf 1}(r\t p) +U(Wp)
\ENN
\EQQ
v  =  \left[\begin{array}{cc}
	{\bf 1}& U\end{array}\right] \left[\begin{array}{c}
	r\t v\\
	Wv
\end{array}\right]= {\bf 1}(r\t v) +U(Wv),
\ENN
one has
\EQQ  \lim_{t\rightarrow \infty} p(t)- p_o(t){\bf 1} = U \lim_{t\rightarrow \infty} Wp(t) =0\\
\lim_{t\rightarrow \infty} v(t)- v_o(t){\bf 1} = U \lim_{t\rightarrow \infty} Wv(t) =0.
\ENN
The proof is thus completed.
\eproof

\bremark The controller (\ref{controller-mu}) consists of two
components. The first component is designed as (\ref{ui-ideal})
for the ideal case with $\xi_i(w_i,v_i)=0$ to achieve consensus.
When the uncertainty $\xi_i(w_i,v_i)$ is taken into account, an
additional adaptive compensator $- \zeta_i(v_i)\mu_i$ with the
update law (\ref{adp-2nd}) is added to the controller.  The
critical advantage of the approach based on Theorem~\ref{thm:mas}
is that the aforementioned two components can be designed
separately. \eremark

\section{Numerical Simulation}
\begin{figure}[htbp]
\begin{centering}
\includegraphics[scale=0.7]{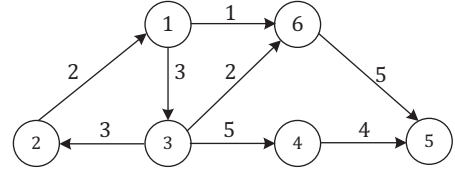}
\par\end{centering}
\caption{The network topology of six agents. \label{fig:topo}}
\end{figure}

We consider a network of $n=6$ agents described by (\ref{agents}) with $\alpha_1=-1$ and $\alpha_2 =0$. 
The nonlinear uncertain terms $\xi_{i}\left(w_{i},v_{i}\right)$'s are given as follows
\EQQ
\xi_{i}(w_{i},v_{i})=\begin{cases}
w_{i}v_{i}^{3}, & i=1,2,3\\
w_{i}, & i=4,5\\
w_{i,1}v_{i}^{3}+w_{i,2}+w_{i,3}v_{i}, & i=6
\end{cases} .
\ENN
%More specifically, $\xi_{i}(w_{i},v_{i})$ for $i=1,2,3$  represents the  nonlinear viscous damping with an unknown parameter $w_i$, $\xi_{i}(w_{i},v_{i})$ for $i=4,5$ is an unknown %constant external force and $\xi_{i}(w_{i},v_{i})$ for $i=6$  is a combination of nonlinear viscous damping, external force and uncertain friction. 
Assume all the unknown parameters are arbitrarily selected within the interval $[-1,1]$.
The network topology is given in Fig.~\ref{fig:topo}
with communication weights marked associated with the edges, also represented by the Laplacian 
\EQQ
L=\left[\begin{array}{cccccc}
2 & -2 & 0 & 0 & 0  & 0\\
0 & 3 & -3 & 0 & 0  & 0\\
-3 & 0 & 3 & 0 & 0  & 0\\
0 & 0 & -5 & 5 & 0  & 0\\
0 & 0 & 0 & -4 & 9  & -5\\
-1 & 0 & -2 & 0 & 0  & 3
\end{array}\right].
\ENN

\begin{figure}[h]
\centering
\includegraphics[scale=0.5]{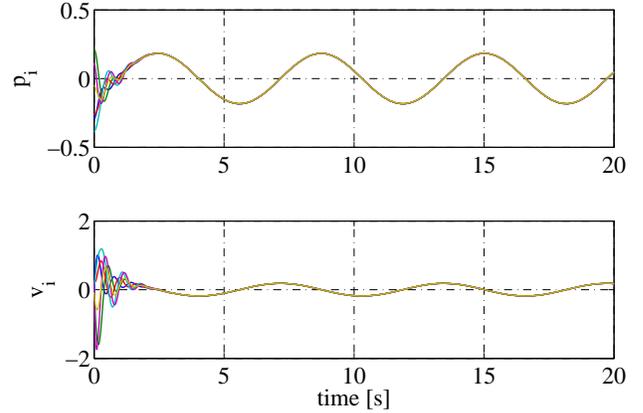}
\par
\caption{Profile of consensus of six agents governed by  (\ref{agents}) with distributed adaptive control. \label{fig:sim}}
\end{figure}

According to Lemma \ref{lemmaA}, we can choose $\gamma_1=15$ and $\gamma_2=1.7$ such that $\bar A$ is Hurwitz and then design the controller (\ref{controller-mu}) and (\ref{adp-2nd}) with $\rho_{i}(v_{i})$ specified as follows
\EQQ
\rho_{i}(v_{i})=\begin{cases}
-\lambda_{i}v_{i}^{4}/4, & i=1,2,3\\
-\lambda_{i}v_{i}, & i=4,5\\
-\lambda_{i}\left[\begin{array}{ccc}
v_{i}^{4}/4 & v_{i} & v_{i}^{2}/2\end{array}\right]\t, & i=6
\end{cases} .
\ENN
The simulation results of the closed-loop system are illustrated in
Fig.~\ref{fig:sim} with $\lambda_i=5$. It is demonstrated that
consensus is asymptotically achieved as concluded by Theorem \ref{thm:mas}.

\section{Conclusion}

In this paper, we have presented a distributed adaptive scheme for an MAS that 
aims to maintain its nominal collective behavior subject to uncertain nonlinearities. The main 
idea is to drive the estimation error to a deliberately designed
manifold in the space of agent states and estimated parameters, which provides 
significant advantages in distributed implementation compared with the traditional 
adaptive law based on gradient of a Lyapunov function. The effectiveness of the new scheme 
has been demonstrated in solving an open asymptotic consensus problem 
for a second-order MAS in a leaderless directed network.  With appropriate design
of the manifold, the scheme is expected to 
handle nonlinearly parameterized uncertainties in the future work. 

\section{Appendix}

{\it Proof of Lemma~\ref{lemmaA}:} Under Assumption~\ref{ass-spanningtree},
all eigenvalues  of $J$ have positive real parts.
Let $P_J \in\mathbb{R}^{(n-1)\times(n-1)}$ be the positive definite
 matrix such that 
\EQQ
P_J J+J\t P_J =I.
\ENN
Let $c$ be a positive constant such that 
\EQQ
P_J<2cI
\ENN
which by Schur complement implies 
\begin{equation}
Q=\left[\begin{array}{cc}
-I & P_J-cI\\
P_J-cI & -c^{2}I
\end{array}\right]<0.\label{eq:Qc}
\end{equation}
By choosing $\gamma_{2}=c\gamma_{1}$
and a sufficiently large $\gamma_{1}>0$, we will show $\bar{A}$ is Hurwitz. Denote 
\EQQ
P=\left[\begin{array}{cc}
\gamma_{1}P_J & P_J\\
P_J & cP_J
\end{array}\right]
\ENN
which is positive definite if $c\gamma_{1}>1$. Note that 
\EQQ
P \bar{A}+\bar{A}\t P=\gamma_{1}Q+Q_{c},
\ENN
where
\EQQ
Q_{c}=\left[\begin{array}{cc}
2\alpha_{1}P_J & (\alpha_{2}+\alpha_{1}c)P_J\\
(\alpha_{2}+\alpha_{1}c)P_J & 2(1+\alpha_{2}c)P_J
\end{array}\right]
\ENN
is a constant matrix. For a sufficiently large $\gamma_{1}$, $P\bar{A}+\bar{A}\t P=\gamma_{1}Q+Q_{c}<0$
due to (\ref{eq:Qc}). Therefore, $\bar{A}$ is Hurwitz.
\eproof

\end{document}